\documentclass[a4paper,11pt]{article}
\pdfoutput=1 

\usepackage{jinstpub} 

\usepackage{lineno}

\title{\boldmath The performance of a new Kuraray wavelength shifting fiber YS-2}

\author[a,b]{I. Alekseev,\note{Corresponding author.}}
\author[b]{M. Danilov}
\author[a]{V. Rusinov,}
\author[a]{E. Samigullin,}
\author[a,b]{D. Svirida,}
\author[a]{E. Tarkovsy}

\affiliation[a]{Alikhanov Institute for Theoretical and Experimental Physics NRC "KI",\\
	B. Cheremushkinskaya 25, Moscow, 117218, Russia}
\affiliation[b]{P.N. Lebedev Physical Institute of the Russian Academy of Sciences,\\
	Leninskiy avenue 53, Moscow, 119991, Russia}

\emailAdd{igor.alekseev@itep.ru}

\abstract{Wavelength shifting fibers are widely used for light
collection from scintillation counters, which allow connection of various
scintillation planes to relatively small photocathodes of photodetectors and
especially tiny photocathodes of silicon photo-multipliers.
In October 2020 Kuraray announced production of a new branch of faster fibers.
We performed a comparison of the new fiber YS-2 to a mature Y-11. The fiber YS-2 
demonstrated decay time nearly two times shorter than that of Y-11: 
$\approx$4.0~ns versus $\approx$7.4~ns. At the same time its light yield 
and attenuation length are as good as of Y-11, which makes YS-2 a good choice 
for timing scintillation detectors.}


\begin{document}
\maketitle
\flushbottom

\section{Introduction}
One of the typical applications of wavelength shifting (WLS) fibers is a light collection from scintillation detectors. 
In usual design a WLS fiber is surrounded by a scintillator by placing 
the fiber in a groove or a hole in the scintillator medium or attached
to the edge of the scintillator plane. In most cases 
this technique is not good for timing measurements. The first problem is in geometry, producing
large variations of light trajectories. But the other problem is in the time distribution
of photon emission by the WLS fiber. Nevertheless there are many cases when even some moderate
time resolution is important. Examples are scintillation counters of CHOD hodoscope of the NA62 
experiment \cite{NA62}, where, in the presence of a very intense beam, the time resolution is important to
suppress overlapping events, a new CMD-3 Time-of-Flight system \cite{CMD-3}, 
or new scintillation detectors for the DANSS experiment upgrade \cite{ICPPA_Dima},
where the time difference between signals from both ends of the scintillation strip 
allows reconstructing the coordinate. Kuraray announcement of the new line of WLS
fibers with shorter decay time looks very promising \cite{YS2}. We report here the results of 
the comparison of 1~mm Kuraray fibers: YS-2MJ Multi (YS-2) and Y-11(200)M Multi (Y-11).

\section{\label{sec:decay}Decay time measurement}
A single photo-electron mode was used to measure the decay time. After an instant excitation of the WLS
fiber at $t=0$ the probability to emit a photon at some time $t$ will follow an exponential law:
\begin{equation*}
N(t) = \left\{
    {\begin{array}{ll}
        0               & \; ,\; \mathrm{when} \; t < 0   \\
        N_0 e^{-t/\tau} & \; ,\; \mathrm{when} \; t \ge 0 \\
    \end{array}} \right. \; ,
\end{equation*}
where $N_0$ is the initial emission intensity and $\tau$ is the decay time. In case of some time measurement 
uncertainty, which follows Gaussian distribution, the probability will be given by a convolution
of an exponential decay and this Gaussian function. The resulting distribution is given by an equation:
\begin{equation}
N(t)=C \left( 1 + \mathrm{erf}(\frac{t - t_0 - \sigma^2/\tau}{\sqrt{2}}) \right) e^{-(t-t_0)/\tau} \; ,
\label{eq:erf}
\end{equation}
where $C$ is the normalization coefficient, $t_0$ is the start time, including light in the fiber 
propagation delay, $\sigma$ is the system time resolution, $\tau$ is the decay time and
$\mathrm{erf}()$ is the Gauss error function \cite{Handbook}. We use short laser pulses (less than 85 ps FWHM) and
select events with only a single photon observed. To extract the decay time the time distribution 
of these events was fit by a sum of formula~(\ref{eq:erf}) and time-independent SiPM noise.

\begin{figure}[htbp]
\includegraphics[width=\textwidth]{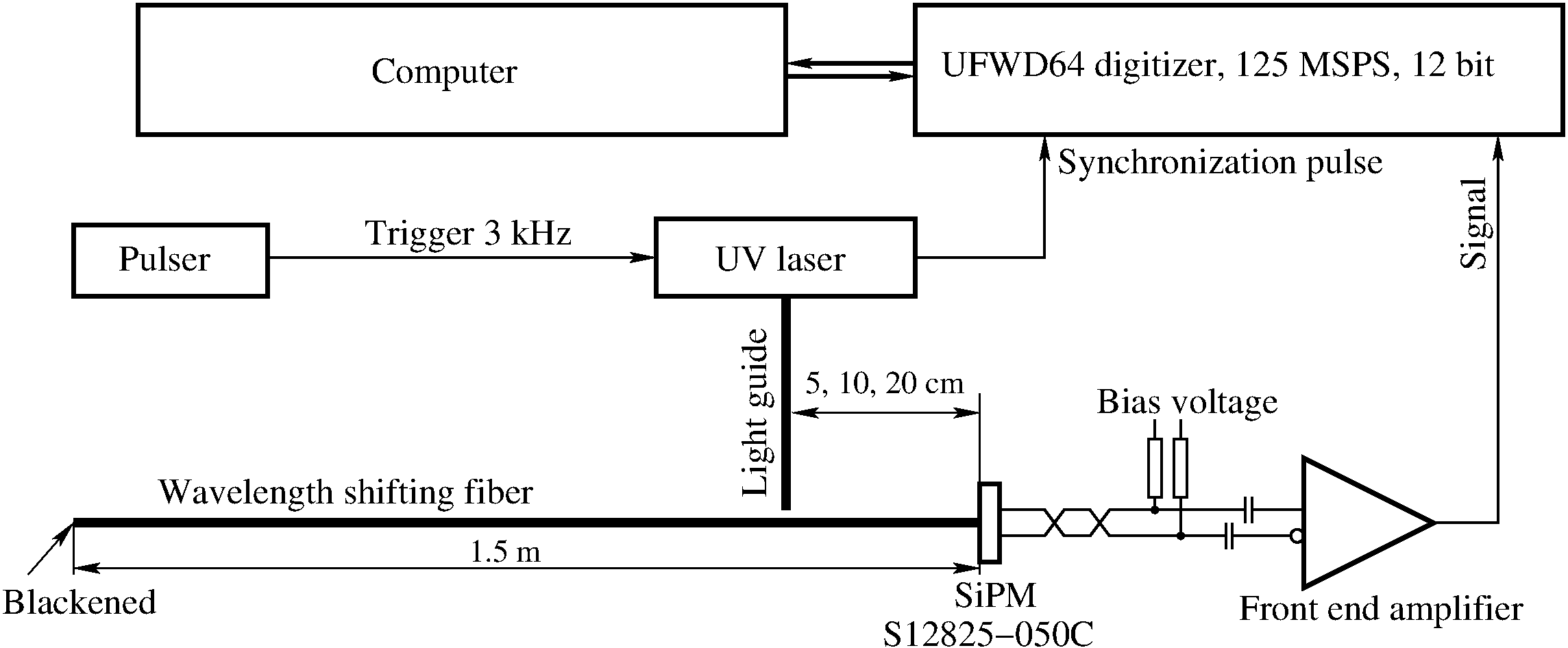}
\caption{\label{fig:setup} A schematic layout of decay time measurement.}
\end{figure}

\begin{figure}[htbp]
\begin{minipage}{0.48\textwidth}
\includegraphics[width=\textwidth,height=0.35\textheight]{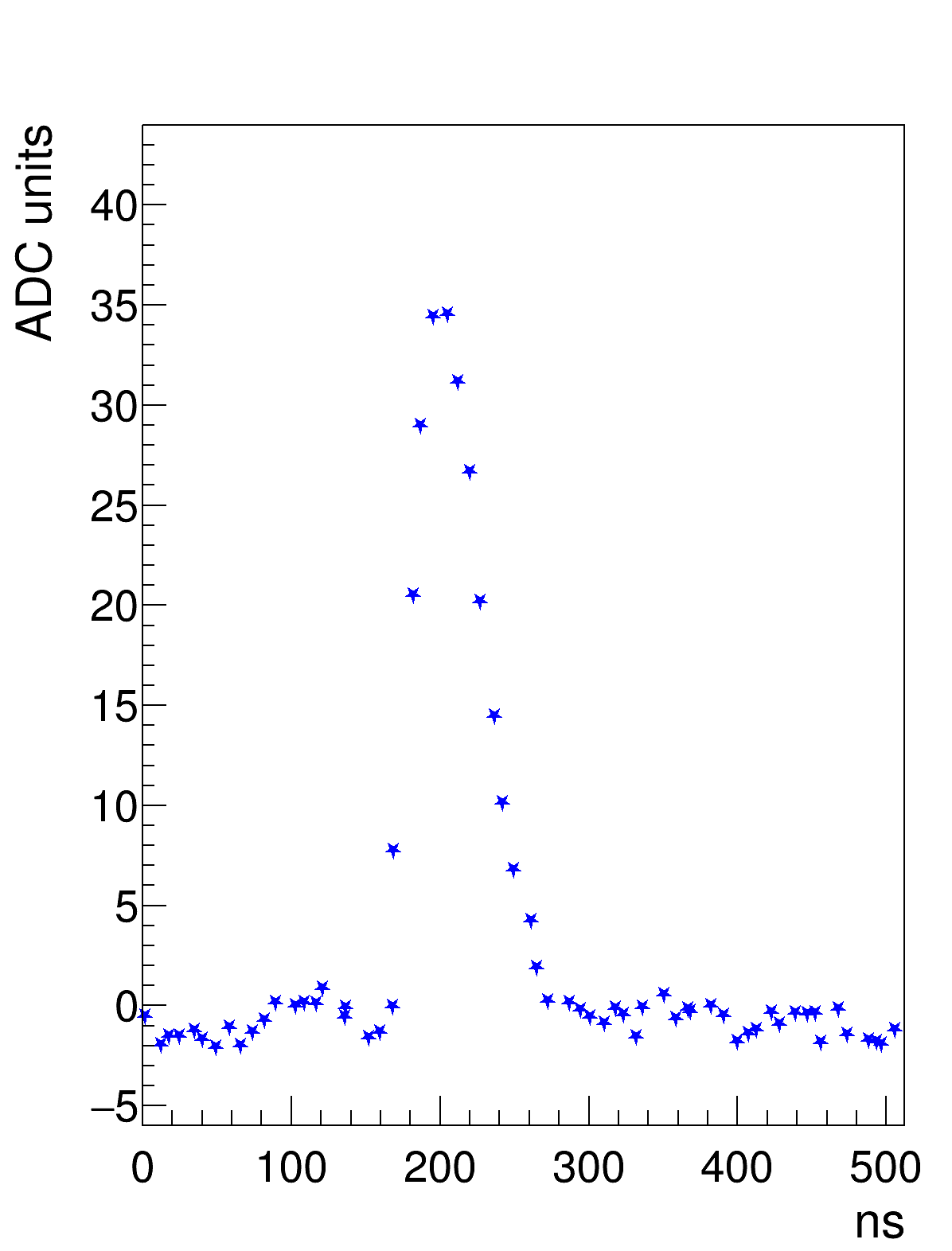}
\caption{\label{fig:signal}Digitized single pixel waveform.}
\end{minipage}
\hspace{0.02\textwidth}
\begin{minipage}{0.48\textwidth}
\includegraphics[width=\textwidth,height=0.35\textheight]{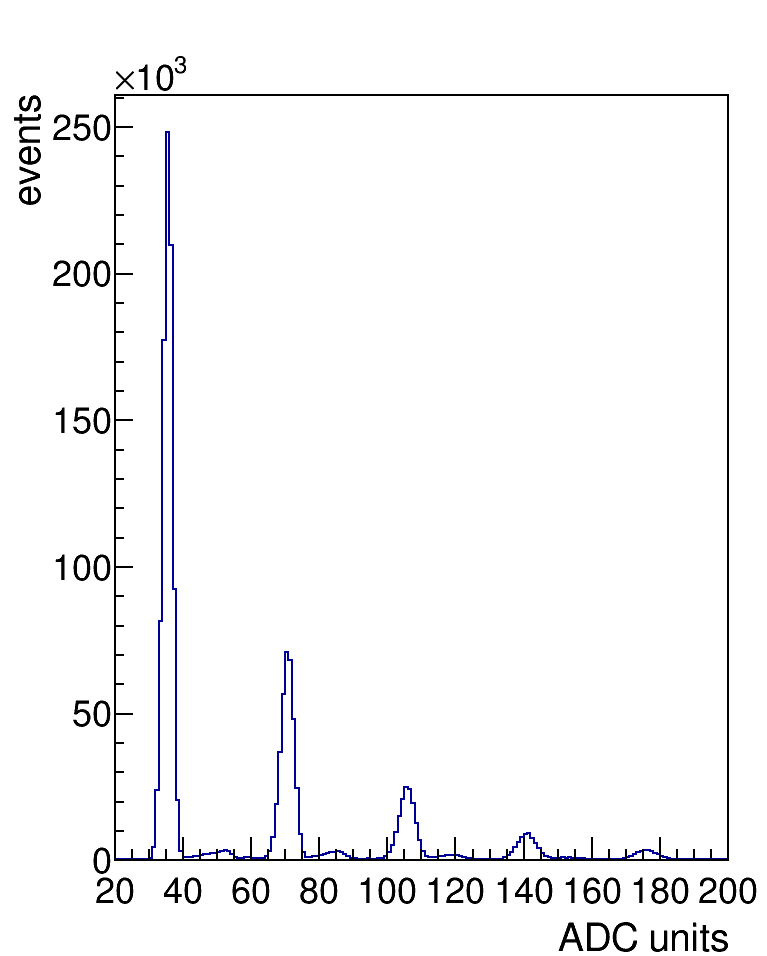}
\caption{\label{fig:ampl}Signal amplitude distribution.}
\end{minipage}
\end{figure}

\begin{figure}[htbp]
\begin{minipage}{0.48\textwidth}
\includegraphics[width=\textwidth]{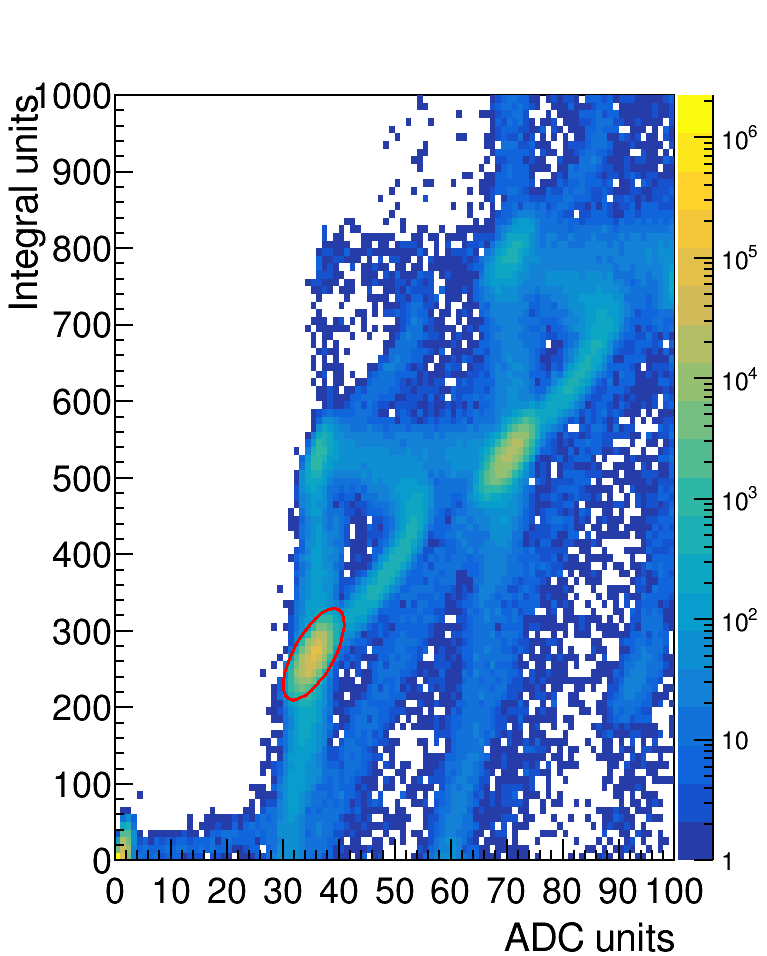}
\caption{\label{fig:2d}Distribution of the pulse charge (integral) over its amplitude. The ellipse
shows the cut, selecting single pixel events.}
\end{minipage}
\hspace{0.02\textwidth}
\begin{minipage}{0.48\textwidth}
\includegraphics[width=\textwidth]{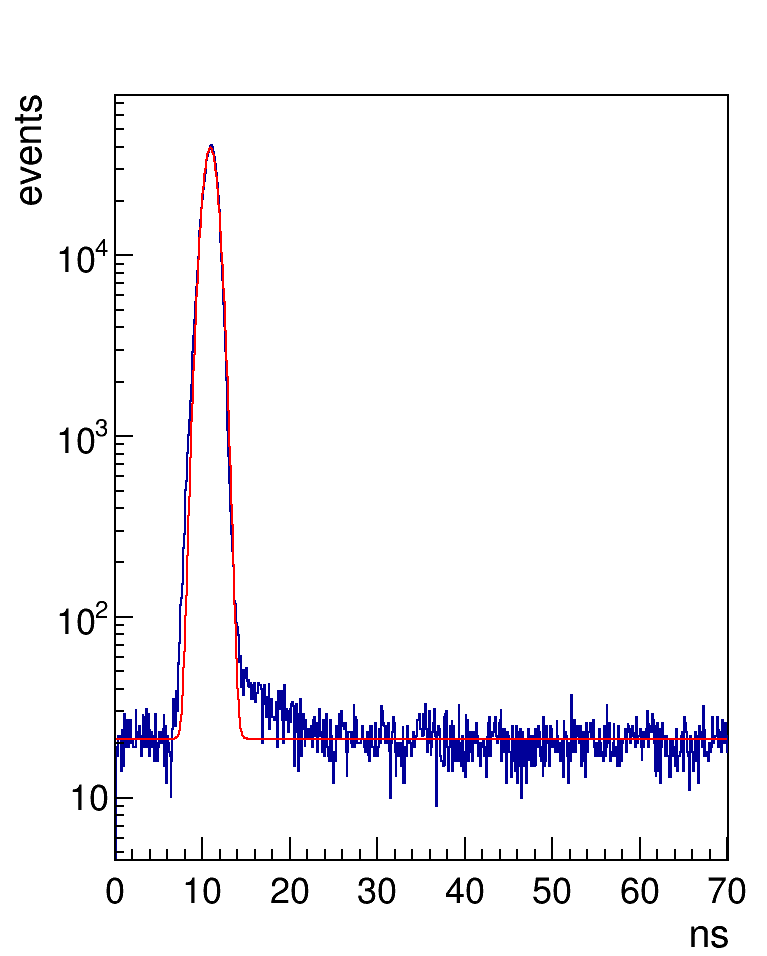}
\caption{\label{fig:SiPM}Single pixel time distribution for direct SiPM illumination. The curve represents
a fit by Gaussian function and uniform background.}
\end{minipage}
\end{figure}

A schematic layout of the decay time measurement is shown in figure~\ref{fig:setup}. A 1.5~m length
piece of tested fiber was polished on one end and connected to HAMAMATSU S12825-050C silicon photo-multiplier (SiPM).
The other end of the fiber was painted black to avoid reflection. An open end of a fiber-optic light
guide was placed a few millimeters from the tested fiber at three different distances 5, 10, and 20~cm from its SiPM end. 
The light guide delivers very short pulses from UV picosecond laser VisUV-266-355-532 by PicoQuant GmbH.
An output with a 355~nm wavelength was used.
The laser was triggered by a pulser with a 3 kHz repetition rate. The signal from SiPM 
was amplified and connected to one input of UVFD64 waveform digitizer module \cite{WFD}. Another input
of the module was used for laser output synchronization pulse, providing the trigger for DAQ and the 
time reference. UWFD64 uses 12 bit 125 MSPS flash ADC to digitize each of its 64 inputs, two of which were
used in this work. An example of a digitized single pixel pulse is shown in figure~\ref{fig:signal}. 
The laser power was adjusted so that only one or no photo-electron was produced in most of the events.
The distribution of events over pulse amplitude is shown in figure~\ref{fig:ampl}. Peaks corresponding
up to five pixels fired are seen. Zero pixel peak is not shown.
For the cleanest selection of single pixel events,
an elliptical cut over two-dimensional distribution charge (integral) versus amplitude is used 
(figure~\ref{fig:2d}). This allows rejecting events with more than one single pixel 
pulses, corresponding to several photo-electrons in one event separated by some time.

\begin{figure}[tb]
\centering 
\includegraphics[width=0.49\textwidth]{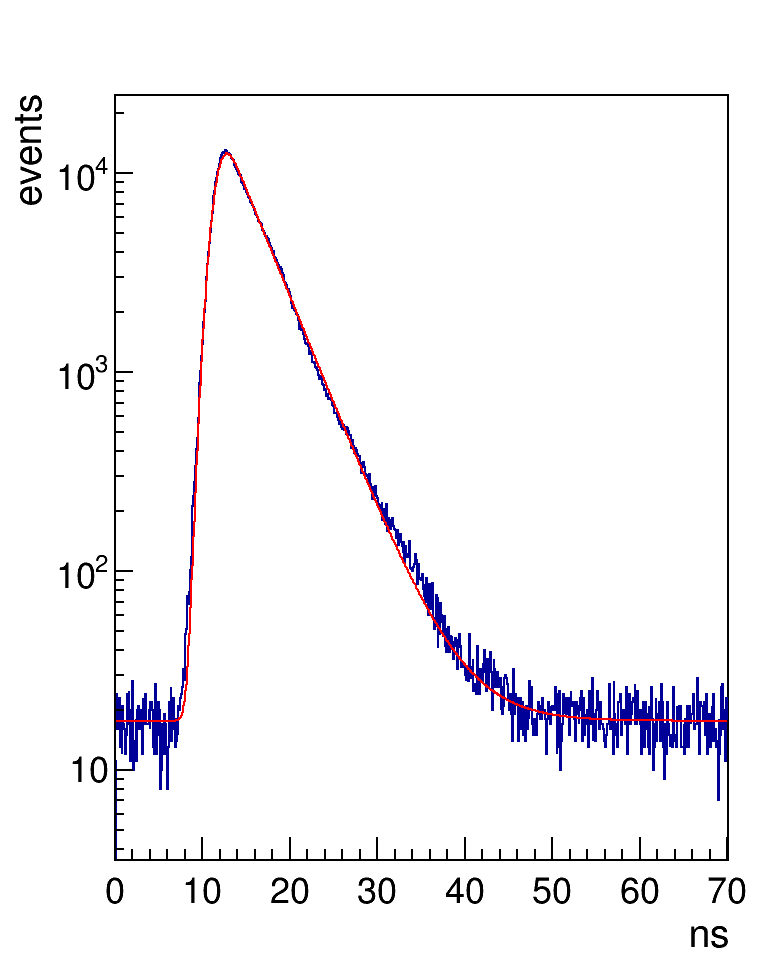}\includegraphics[width=0.49\textwidth]{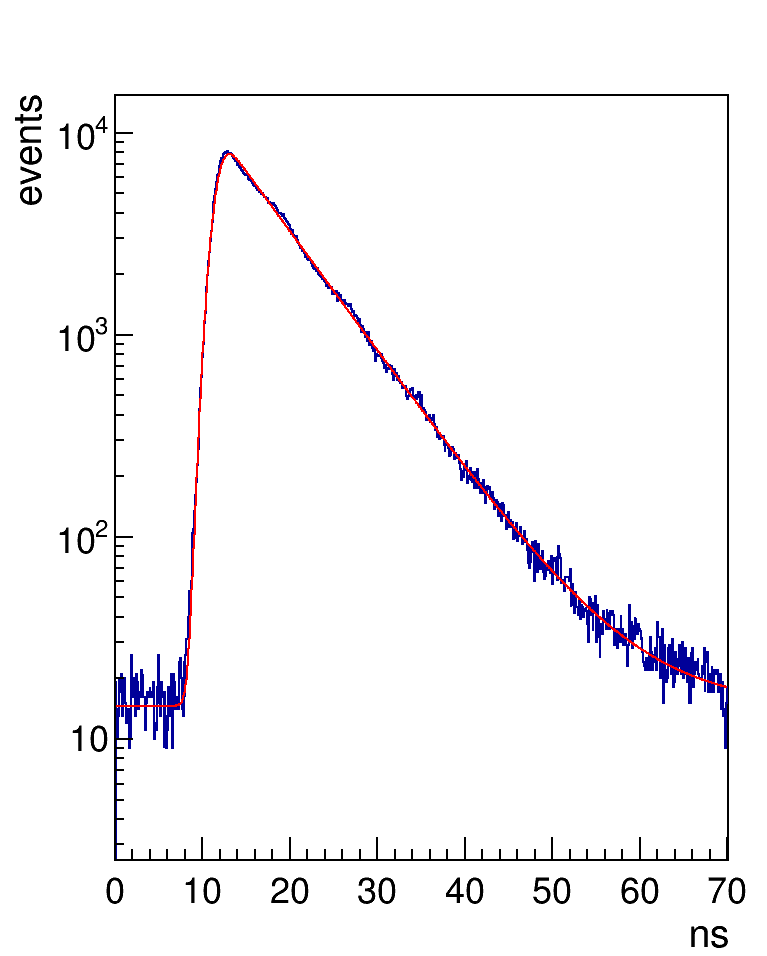}
\caption{\label{fig:YS2Y11}Single pixel time distribution for fibers YS-2 (left) and Y-11 (right). 
The curve represents a fit by the function (\ref{eq:erf}) and uniform background.}
\end{figure}

The system time resolution was measured in a dedicated
experiment, where a laser light guide was directed to the SiPM without a WLS fiber. 
The time distribution of single pixel events was fitted by a sum of Gaussian and
uniform background (see figure~\ref{fig:SiPM}). The measured $\sigma = 0.8019 \pm 0.0007$~ns
was fixed for further fits of the fiber measurements. The decay time of both YS-2 and Y-11 WLS fibers 
was measured at 3 distances from the laser beam spot to the SiPM. 
No dependence on the distance was observed. 
Time distributions for 10~cm distance are shown in figure~\ref{fig:YS2Y11}. 
The results of the measurements are summarized in table~\ref{tab:res}.

\begin{table}[htbp]
\caption{\label{tab:res}The measured decay time $\tau$ in nanoseconds. Only statistical errors are given
in 5, 10 and 20 cm columns. The last column presents the values provided by the manufacturer \cite{YS2}.}
\centering
\begin{tabular}{|l|ccc|c|c|}
\hline
Fiber    & 5 cm              & 10 cm             & 20 cm             & Average $\pm$ RMS & Kuraray datasheet\\
\hline
YS-2     & $3.949 \pm 0.006$ & $4.005 \pm 0.006$ & $4.003 \pm 0.005$ & $3.99 \pm 0.03$   & 3.2 \\
Y-11     & $7.293 \pm 0.010$ & $7.295 \pm 0.010$ & $7.522 \pm 0.010$ & $7.37 \pm 0.11$   & 6.9 \\
\hline
\end{tabular}
\end{table}

We collected about 2 million events for each configuration out of which 0.8-0.9 million satisfy 
single pixel criteria, so the statistics in each point are very high and statistical errors are small.
Values of the decay time $\tau$ are stable to the fit range and Gaussian width $\sigma$ variation.
An estimation of the systematic error could be estimated from the result variation
at different distances. RMS of the values at 3 distances was calculated and indicated as an
error in the last but one column of the table~\ref{tab:res}.

\section{Light output measurement with cosmic rays}
\begin{figure}[htbp]
\centering\includegraphics[width=0.5\textwidth]{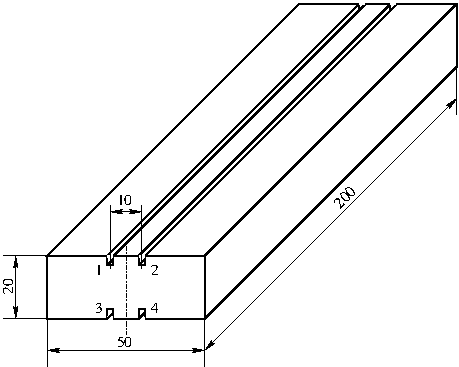}
\caption{\label{fig:counter}Layout of the test counter with 4 grooves. Grooves numbers are also shown.}
\end{figure}

An important question for WLS fiber is the light output in the real detector
geometry. For the comparison of YS-2 and Y-11 a sample scintillation counter 
with 4 grooves was manufactured from polystyrene (see figure~\ref{fig:counter}). 
Fibers were cut at 23 cm and polished on both ends. One end of a fiber was 
connected to SiPM and the other end was left open. Four fibers 
of one type were put into the grooves without any glue and exposed to cosmic
rays. Then fibers were exchanged keeping SiPM and grooves correspondence and
another exposition taken. The results are presented in table~\ref{tab:cosmres}.
Individual values have a statistical error of about 2\% not shown in the table.
All 4 fibers are in the same conditions, so their variation is used for
error estimation, which is shown in the last column together with average values.
This simple test demonstrated that the light yield of YS-2 is at least as good as 
of Y-11. 

\begin{table}[!h]
\caption{\label{tab:cosmres}Median light output for cosmic particles crossing the test scintillator
counter. Values are given for each fiber.}
\centering
\begin{tabular}{|l|cccc|c|}
\hline
Groove      & 1    & 2    & 3    & 4    & Average $\pm$ RMS \\
\hline
YS-2, ph.e. & 30.7 & 24.6 & 28.2 & 25.5 & $27.3 \pm 2.4$ \\
Y-11, ph.e. & 24.8 & 25.5 & 28.6 & 22.7 & $25.4 \pm 2.1$ \\
\hline
\end{tabular}
\end{table}

\section{Longitudinal attenuation measurement}
\begin{figure}[htbp]
\centering\includegraphics[width=0.5\textwidth]{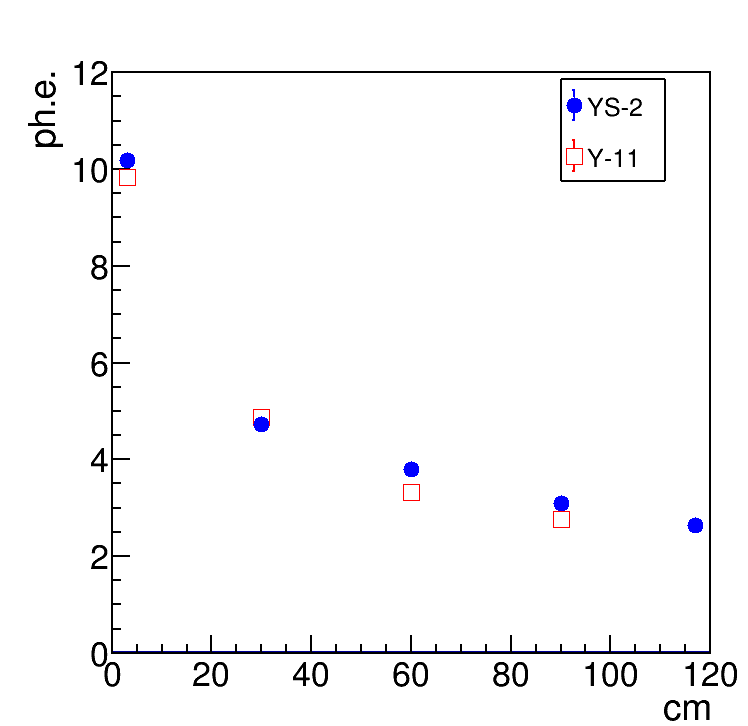}
\caption{\label{fig:long}Signal longitudinal attenuation in YS-2 (full circles) and Y-11 fibers (open squares).}
\end{figure}

Measurement of light attenuation was performed with the help of $^{90}$Sr beta source
placed above a rectangular piece of the scintillator ($30 \times 30 \times 3$~mm$^3$) with
a hole drilled along its 30~mm side. A scintillation counter was placed below the scintillator with the hole
and allowed to trig on the electrons passing through it. The set of the source and scintillators may slide 
along the tested fiber for measurements at various distances.
1.2~m length pieces were used, polished on both ends with SiPM attached to one end.
Median light output in photo-electrons is shown in figure~\ref{fig:long} as a function of the distance from 
the readout end. Longitudinal attenuation of YS-2 looks to be as small as of Y-11. Our YS-2 samples 
were relatively short, so we cannot compare our attenuation length value for YS-2 ($1.45 \pm 0.07$~m), 
calculated using the data between 0.3 and 1.2~m, with manufacture's value of 3.5~m,
corresponding to distances $1.0 - 3.0$~m \cite{KURARAY}. 

\section{Conclusions}
Timing response of scintillation counters with WLS fibers is already studied relatively
well \cite{WOJCIK,LHCB,IHEP} and many more. The decay time measured for Y-11 fiber 
is in an agreement with the previous measurement \cite{LHCB}. In that work a fit
of large (many photons) signal shape was performed and the result was $8.8 \pm 1.5$~ns for Y-11(M200)
and $7.3 \pm 1.1$~ns for Y-11(MS250). This time is essentially larger than that of BCF-92 by Saint-Gobain \cite{BCF92}
($2.4 \pm 0.4$~ns \cite{LHCB}). But  BCF-92 demonstrates significantly lower light yield --
the average ratio for various scintillation materials and the same detector geometry between
Y-11 and BCF-92 is $1.85 \pm 0.07$ \cite{IHEP}. For the new Kuraray WLS YS-2 the light output and attenuation length 
are as good as of Y-11. The decay time of YS-2 is nearly two times shorter. This makes YS-2 a good choice for 
timing measurements, leaving well behind Y-11 with its comparatively large decay time and BCF-92 with its comparatively
low light yield.

\acknowledgments
This work is supported by the Ministry of Science and Higher Education of the Russian
Federation under Contract No. 075-15-2020-778.


\end{document}